\documentclass[final]{aipproc}
\usepackage{graphicx,amsmath,amsfonts,amssymb,mathrsfs}

\layoutstyle{8x11double}

\newcommand{\eg}{{\it e.g.}}
\newcommand{\cf}{{\it cf.}}

\newcommand{\fig}{Figure}

\newcommand{\Ref}{Ref.}
\newcommand{\Refs}{Refs.}

\newcommand{\stheta}{\sin^22\theta_{13}}
\newcommand{\deltacp}{\delta_{\mathrm{CP}}}
\newcommand{\ldm}{\Delta m_{31}^2}
\newcommand{\sdm}{\Delta m_{21}^2}

\newcommand{\figu}[1]{\fig~\ref{fig:#1}}

\newcommand{\bi}{\begin{itemize}}
\newcommand{\ei}{\end{itemize}}

\begin{document}

\title{Optimization of a Neutrino Factory:\\ Discovery Machine versus Precision Instrument}

\classification{14.60.Pq
%<Replace this text with PACS numbers; choose from this list:
%                \texttt{http://www.aip..org/pacs/index.html}>
}
\keywords      {Neutrino Oscillations, Neutrino Factory, Leptonic CP violation}

\author{Walter Winter}{
  address={Institut f{\"u}r theoretische Physik und Astrophysik, Universit{\"a}t W{\"u}rzburg \\
Am Hubland, D-97074 W{\"u}rzburg, Germany}
}

\begin{abstract}
We discuss the optimization of a neutrino factory experiment for the purpose of 
$\stheta$, mass hierarchy, and CP violation discoveries. This includes a review of
possible optimization strategies, as well as an application of these to different $\stheta$
regions. 
\end{abstract}

\maketitle

%%%%%%%%%%%%%%%%%%%%%%%%%%%%%%%%%%%%%%%%%%%%
%% MAINMATTER
%%%%%%%%%%%%%%%%%%%%%%%%%%%%%%%%%%%%%%%%%%%%

\section{Introduction}

The neutrino oscillation parameters are an important component
to construct a theory for lepton masses and mixings. In lepton mass models, any observables 
describing deviations from potential symmetries turn out to be good performance indicators,
which can be used to test the model parameter space. For example, the yet unknown value of $\stheta$ and
the neutrino mass hierarchy are indicative for certain classes of models found
in the literature, such as flavor symmetries or grand unified theories~\cite{Albright:2006cw}. Similarly, 
deviations from maximal atmospheric
mixing~\cite{Antusch:2004yx} may describe deviations from a $\nu_\mu$-$\nu_\tau$ symmetry, and the
phenomenological relationship $\theta_{12} + \theta_C \simeq \pi/4$ (``quark-lepton
complementarity'') may be a performance indicator for quark-lepton unification~\cite{Smirnov:2004ju}.
In addition, there may be a connection between $\deltacp$ and leptogenesis, which 
motivates the search for leptonic CP violation. 

Three-flavor neutrino oscillations can be described by six parameters:
The solar parameters $\sdm$ and $\theta_{12}$, the atmospheric parameters $\ldm$ and
$\theta_{23}$, the small mixing angle $\theta_{13}$, and the leptonic Dirac CP phase
$\deltacp$. At this time, we know the solar and atmospheric oscillation parameters very well,
but we do not know the sign of $\ldm$ (neutrino mass hierarchy), the value of $\theta_{13}$
(we only have an upper bound for), and the leptonic CP phase (see, \eg, \Ref~\cite{Schwetz:2006dh}).
In this talk, we therefore focus on the discoveries of (nonzero) $\stheta$, the neutrino mass hierarchy, and leptonic CP violation by means of a neutrino factory~\cite{Geer:1997iz,Albright:2000xi,Apollonio:2002en} producing neutrinos by
muon decays in straight sections of a storage ring.
The unprecedented reach and accuracy of a neutrino factory, and the 
broad scope of physics that can be explored  has been discussed in detail in 
several studies (see, \eg, \Refs~\cite{Barger:1999jj,Barger:1999fs,Barger:2000cp,Cervera:2000kp,Burguet-Castell:2001ez,%
Burguet-Castell:2002qx,Huber:2002mx,Huber:2006wb}). 
In particular, recently, the resolution of various degeneracies~\cite{Burguet-Castell:2001ez,Fogli:1996pv,Minakata:2001qm,Barger:2001yr} 
has been an important topic.

\begin{figure}[t]
\includegraphics[width=\columnwidth]{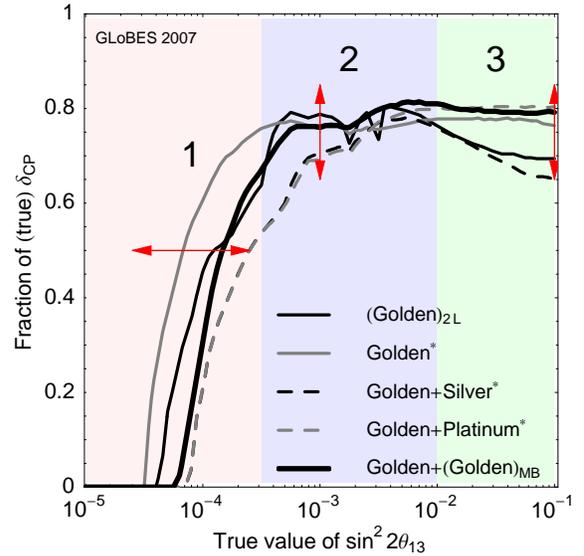}
\caption{\label{fig:cpf} Fraction of $\deltacp$, for which CP violation can be discovered,
as function of the true $\deltacp$ for various different options ($3 \sigma$). The three different
optimization regions discussed in this talk are marked by different shadings and arrows.
Similar figure as in \Ref~\cite{Huber:2006wb}.}
\end{figure}

For the optimization of a neutrino factory, different $\stheta$ ranges are relevant which are illustrated in \figu{cpf} for the CP violation measurement. In this figure, we show the fraction of all possible (true) values of $\deltacp$ for which one could establish CP violation as a function of the (true) $\stheta$.
If we did not find $\stheta>0$ in the next generation of experiments, it might 
the primary objective to optimize for as small $\stheta$ as possible. For CP violation, that strategy
is illustrated in the shaded region~1 in \figu{cpf}, where one would optimize along the respective arrow.
This case is mainly limited by statistics and systematics. If, however, $\stheta$ turned out to be large, then one would optimize the fraction of $\deltacp$, for example, along the arrow in region~3.
In this case, the main limitations are correlations among the oscillation parameters, such as with the matter density uncertainties (see, \eg, \Ref~\cite{Ohlsson:2003ip}). In addition, the relatively high energy threshold above the interesting oscillation pattern affects the performance~\cite{Huber:2002mx}. If, however, $\stheta$ was in region~2, one would optimize the fraction of $\deltacp$. In that region, discrete degeneracies limit the performance.
Since the fraction of $\deltacp$ for which one can measure CP violation is close connected to the precision of $\deltacp$ at the CP conserving values, regions~2 and~3 correspond to an optimization of precision, whereas region~1 represents the discovery limit.

In this talk, we will discuss the optimization for regions~1,~2, and~3 separately. However,
note that the next generation of experiments, such as superbeams and reactor experiments, will 
discover $\stheta$ only if $\stheta \gtrsim 0.01$~\cite{Huber:2004ug,Albrow:2005kw}.
In the discovery case, one will exactly know what $\stheta$ to optimize for.
If on the other hand $\stheta$ is not discovered by these experiments, one will not know if $\stheta$ is in region~1 or~2. Therefore, if an decision has to be made at this point, a potentially contradictive optimization outcome between regions~1 and~2 will be very unfortunate. We will come back to this discussion in the conclusions.

\section{Optimization Options}

Possible optimization options for a neutrino factory include the baseline(s),
the muon energy, the combination of different oscillation channels,
an optimization of the detector, or a potential luminosity increase\footnote{
In the following, we will refer to double luminosity by ``2L''.}.
In addition, one could think of combinations with different experiment classes,
such as superbeams~\cite{Burguet-Castell:2002qx,Huber:2007uj}.

As potentially interesting appearance channels, we have the 
$\nu_e \rightarrow \nu_\tau$ (``silver'')~\cite{Donini:2002rm,Autiero:2003fu}
and $\nu_\mu \rightarrow \nu_e$ (``platinum'') channels~\cite{Bueno:2001jd}
in addition to the standard $\nu_e \rightarrow \nu_\mu$ (``golden'') channel~\cite{Cervera:2000kp,Burguet-Castell:2001ez}
using a magnetized iron calorimeter (MID).\footnote{The different channels can also
be operated with antineutrinos.}
For the silver channel, one typically assumes an OPERA-like emulsion cloud chamber (ECC) as
a detector. Note that the silver 
channel requires relatively high muon energies because of the $\tau$
production threshold. In addition, one usually assumes that only the leptonic decay
modes of the $\tau$ can be observed. For this talk, we will adopt the optimistic point 
of view that the hadronic decay modes can be observed as well, and that one can
built a $10 \, \mathrm{kt}$ detector called ``Silver*''. For the platinum channel,
the problem is that the produced electrons start producing showers very quickly, which
makes charge identification at high energies practically impossible in an iron calorimeter.
Again, we adopt the optimistic point of view that a detector technology without
that problem can be found (such as liquid argon), and call the detector ``Platinum*''
(for details, see \Ref~\cite{Huber:2006wb}).
Note that the platinum channel is the T-inverted channel to the ``golden'' $\nu_e \rightarrow \nu_\mu$
channel. In principle, it allows for a CP violation measurement without having to disentangle the
$\deltacp$ effects from the matter effects (see, \eg, \Ref~\cite{Akhmedov:2004ny}).

As for the detector optimization, a lower energy threshold may be achieved 
for the price of higher backgrounds~\cite{Anselmo}. This lower energy threshold
would allow for lower muon energies because the energy range, where the main oscillation effects
take place, can be covered with a better efficiency. For example, for very large 
$\stheta$, it allows for muon energies as low as $E_\mu \gtrsim 4 \, \mathrm{GeV}$,
and leads to a re-optimization of the baseline~\cite{Huber:2007uj,Geer:2007kn,Bross:2007ts}.
This relatively new idea will be referred to as a ``low energy neutrino factory''. The
the corresponding detector with a lower threshold will be called ``Golden*'' in the following.

An interesting particular baseline option is the ``magic baseline'' (MB). It turns out that,
two second order in $\sin 2 \theta_{13}$ and $\alpha=\sdm/\ldm$, the dependence on
$\deltacp$ disappears at this baseline $L \simeq 7 \, 500 \, \mathrm{km}$ independent of
the oscillation parameter values and neutrino energy~\cite{Huber:2003ak}. This effect can be
used for a degeneracy-free measurement of $\stheta$ and the mass hierarchy.
The exact baseline depends on specifics of the matter density profile. However, in combination with a short baseline, the matter profile dependence is negligible as long as the second baseline is long enough ($7 \, 000 \, \mathrm{km} \lesssim L \lesssim 9 \, 000 \, \mathrm{km}$)~\cite{Gandhi:2006gu}.

In some cases, we will compare the results with the $\gamma=350$ beta beam from \Ref~\cite{BurguetCastell:2005pa}, called ``Beta beam'' in the following.

\section{Optimization Regions}

\begin{figure}[t!]
\includegraphics[width=\columnwidth]{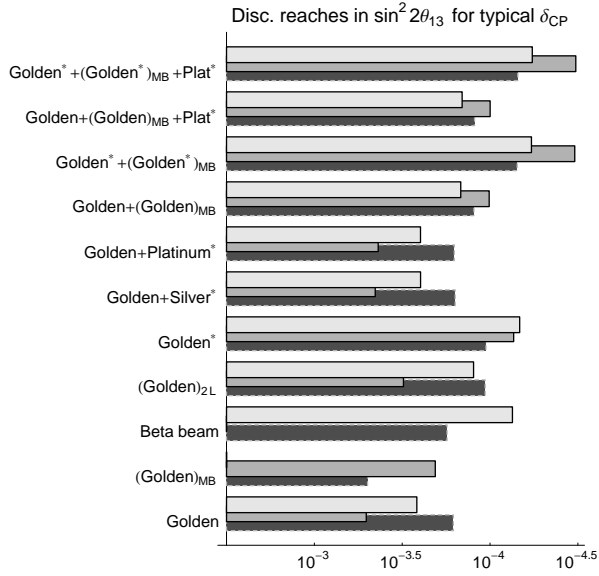}
\caption{\label{fig:disc3} Comparison of options for CP violation (light bars, $3 \sigma$), mass hierarchy (medium gray bars, $3 \sigma$), and $\stheta$ discovery reaches (dark bars, $5 \sigma$). The figure shows the $\stheta$ reach for a CP fraction of 0.5 (``typical $\deltacp$''). All bars Figure taken from \Ref~\cite{Huber:2006wb}.}
\end{figure}

In the following, we discuss the optimization along the arrows in \figu{cpf} for the three different
regions in that figure.

{\bf Region~1} (discovery for smallest $\stheta$): The simultaneous baseline and muon energy optimization
was discussed in \Ref~\cite{Huber:2006wb} (see also \Refs~\cite{Barger:1999fs,Cervera:2000kp,Burguet-Castell:2001ez,Freund:2001ui}). For the resolution of
degeneracies and the mass hierarchy measurement, the magic baseline turns out to be the best choice, and
for the CP violation measurement, a shorter baseline in a relatively broad window $L \sim 2 \, 000 - 5 \, 000 \, \mathrm{km}$. Therefore, the combination of these two baselines is the currently discussed baseline setup of a neutrino factory~\cite{ISS}. \figu{disc3} summarizes the results for all three discovery reaches
and for many different upgrade options. Note that the figure corresponds to the performance
along the left arrow in \figu{cpf}.
From \figu{disc3}, the magic baseline and the improved detection system are the key elements to improve the $\stheta$ reach. The Golden* detector allows for muon energies as low as $E_\mu \simeq 20 \, \mathrm{GeV}$ at the discovery limit without loss of physics potential~\cite{Huber:2006wb}.

\begin{figure*}[t!]
\includegraphics[width=\textwidth]{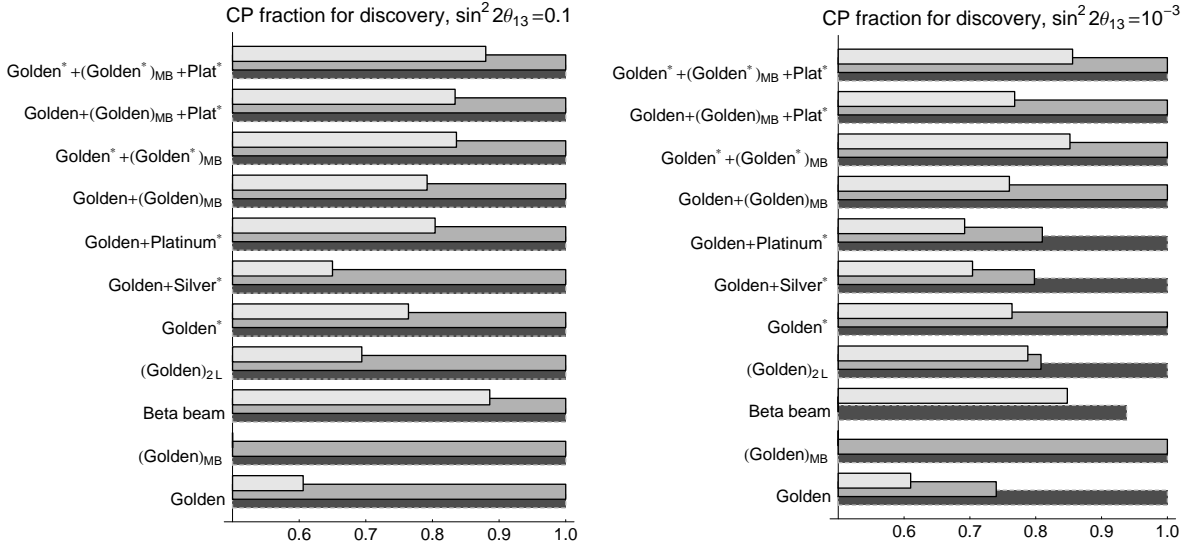}
\caption{\label{fig:disc12} Comparison of options for CP violation (light bars, $3 \sigma$), mass hierarchy (medium gray bars, $3 \sigma$), and $\stheta$ discovery reaches (dark bars, $5 \sigma$). The figures show the fraction of $\deltacp$
(``CP fraction'') for a given true $\stheta$ (see figure captions).  Figure taken from \Ref~\cite{Huber:2006wb}.}
\end{figure*}

{\bf Region~2} (precision for intermediate $\stheta$): A number of  different optimization options in this case are compared in \figu{disc12}, right panel. The conclusions are similar to the ones in region~1: magic baseline and improved detection system would clearly help. Compared to region~1, however, the $\stheta$ discovery turns out to be not a problem for almost any of the discussed options, and the discussion centers around the mass hierarchy and CP violation discoveries.
One can also read off this figure that the silver and platinum channels do have some degeneracy resolution potential. This, however, does not lead to a substantial physics potential increase beyond the use of magic baseline or improved detector anymore.
Note that the usefulness of the magic baseline as a risk minimizer can be also seen in the $\deltacp$~\cite{Huber:2004gg} and $\stheta$ precision~\cite{Gandhi:2006gu} measurements.
For example, a $\deltacp$ precision of about 10 degrees can be achieved ($1\sigma$) independent of the true $\deltacp$ with a combination of two baselines. Such a precision comparable to the magnitude of the Cabibbo angle can, for example, be motivated in quark-lepton complementarity based models~\cite{Winter:2007yi}.

{\bf Region~3} (precision for large $\stheta$): In this region, the value of $\stheta$ will be
likely discovered soon, and the $\stheta$ discovery is not relevant for the optimization
anymore. In addition, as one can see in \figu{disc12}, left panel, the mass hierarchy can be easily
determined for all possible values of $\deltacp$. Therefore, the optimization is determined by CP violation. 
In principle, there are two options for a neutrino factory discussed in the literature: A high energy option ($E_\mu \gtrsim 20-50 \, \mathrm{GeV}$), and a
low energy option ($E_\mu \sim 5 \, \mathrm{GeV}$) if there are substantial improvements in the detection threshold (such as a different detector technology can be used).

Any optimization discussion of a neutrino factory for large $\stheta$ should take into account that there are competing experiments as well. For example, superbeam upgrades do have a similar physics potential and can be optimized, too (see, \eg, \Refs~\cite{Barger:2006kp,Barger:2007jq} for an optimization discussion of different options). In addition, several higher gamma beta beam options have been discussed in the literature~\cite{Burguet-Castell:2003vv,Huber:2005jk,BurguetCastell:2005pa,Agarwalla:2006vf,Agarwalla:2007vz}, which may even be competitive for smaller $\stheta$. Therefore, we will use several representatives of these experiment classes for comparison.

For the optimization of the high neutrino energy option, several possibilities are summarized in \figu{disc12}, left panel. In addition to magic baseline and improved detector, the platinum channel helps in this region. However, the $\gamma=350$ beta beam shown for comparison has a very competitive physics reach. 
This comparison, of course, depends on the useful parent decays of the two different experiment classes, which are very difficult to compare until there are dedicated cost studies for both options. In addition to the possibilities discussed in \figu{disc12}, improved knowledge on the matter density profile helps for large $\stheta$. However, if different channels and baselines are combined, this knowledge becomes obsolete~\cite{Huber:2006wb}, and, in fact, one can measure the average lower mantle density of the Earth at the level of 0.5\% ($1\sigma$) with the very long baseline~\cite{Gandhi:2006gu,Winter:2005we,Minakata:2006am}.

\begin{figure*}[t!]
\includegraphics[width=0.7\textwidth]{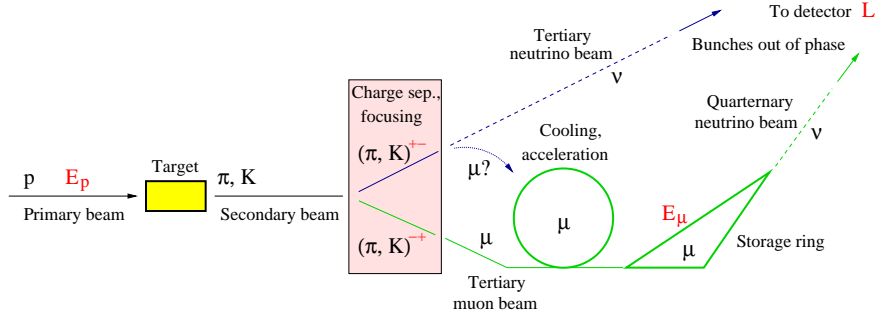}
\caption{\label{fig:schematics} Schematics of a NF-SB (not to scale). Our
degrees of freedom are given by red/gray labels. Note that we adopt the conservative
point of view that only half the muons can be collected for the NF. Figure from \Ref~\cite{Huber:2007uj}.}
\end{figure*}

If the detection system can be improved and effort (in terms of muon energy) matters, the muon energy can be significantly reduced~\cite{Huber:2007uj,Geer:2007kn,Bross:2007ts}. Several low energy neutrino factory options, as well as their optimization have been discussed in \Ref~\cite{Huber:2007uj}.
In particular, it may be an option to combine a neutrino factory with a superbeam directed to the same
detector at the same baseline, which we call neutrino factory superbeam (NF-SB)~\cite{Huber:2007uj}. This superbeam may even be produced in the same decay chain as the neutrino factory beam, or it may be produced by using two targets. A schematics for such an experiment can be found in \figu{schematics}. The main idea is to have a platinum-like $\nu_\mu \rightarrow \nu_e$ channel with a spectrum shifted towards lower neutrino energies compared to the platinum spectrum (the platinum spectrum is always peaking at higher energies compared to the golden channel because of the muon decay kinematics). In addition, out-of-phase bunches will not require charge identification of electrons at the detector. Since the interesting oscillation pattern tends to appear at lower energies, the absolute performance of a NF-SB is better than for a neutrino factory -- even if the double luminosity is assumed for the neutrino factory alone (or only half of the muons can be collected for the neutrino factory mode). The optimal baseline is about $800$ to $1 \, 500 \, \mathrm{km}$ with $E_\mu \gtrsim 5 \, \mathrm{GeV}$ and $E_{\mathrm{Prot}}=28 \, \mathrm{GeV}$. The NF-SB shows a very competitive behavior in the whole range $0.01 \lesssim  \stheta \lesssim 0.1$, such as compared to a wide band superbeam upgrade using a megaton-size water Cherenkov detector. A similar result could be achieved for a low energy neutrino factory in combination with the platinum channel. That option, however, requires electron charge identification.

\section{Summary and Conclusions}

In the previous discussion, we have optimized the different $\stheta$ regions for $\stheta \lesssim 0.01$ separately (\cf,  \figu{cpf}), because they correspond to different optimization goals: maximum reach in $\stheta$ (discovery limit) versus maximal reach in $\deltacp$ (need for precision). However, it has turned out that the regions~1 and~2 have the same requirements: an improved detection system and a second (very long) baseline turn out to be key components of an optimized neutrino factory for $\stheta \lesssim 0.01$. Since we will, most likely, not know how big $\stheta$ is when we have to make a decision for region~1 or~2, the similar optimization outcome turns out to be very fortunate.

If $\stheta \gtrsim 0.01$, we will have established a nonzero $\stheta$ very soon by the planned reactor and superbeam experiments. In this case, there are two options for a neutrino factory which are very different from the experimental point of view: They require different muon energies, different baselines, and, most importantly, a different detector.  As for the combination with different channels, it remains to be clarified how well the platinum channel (electron neutrino detection with charge identification) can be implemented, and up to which energies. In sorting out the different options, the detector optimization and the test of different detector technologies are the key ingredients, and it should be one of the primary objectives for the coming years.

In this study, we have only focused on the optimization for the unknown neutrino oscillation parameters.
However, different requirements, such as the search for non-standard physics or sensitivity to deviations from maximal atmospheric mixing, may lead to different optimizations. In the future, one will need to study how these different objectives fit together, and if there are any other unexplored experimental approaches.

{\bf Acknowledgments}
 I would like to acknowledge support from the Emmy Noether program of Deutsche Forschungsgemeinschaft.

%\bibliographystyle{aipproc}  
%\bibliography{references}

\end{document}